# Quantum coherence in dissociative electron attachment: isotope effect

Suvasis Swain[1], Akshay Kumar[1], E. Krishnakumar[2], and Vaibhav S. Prabhudesai[1*]

[1] Tata Institute of Fundamental Research, Mumbai 400005 India
[2] Raman Research Institute, Bengaluru 560080 India
*vaibhav@tifr.res.in

**Abstract:**

Dissociative electron attachment (DEA) is one of the processes that shows a strong coupling between the nuclear and electronic degrees of freedom in a molecule. This coupling results in an efficient transformation of the kinetic energy of attaching free electrons into the chemical energy of the compound molecule. A recent discovery of quantum coherence in this process has opened a whole new dimension in its description. On the other hand, the mass variation in isotopes of the constituent atoms has a profound effect on DEA. In quantum coherence observed in DEA, the isotope effect depicts itself in terms of change in the phase and the amplitude of the interfering dissociation paths. Here, we report the quantum coherence observed in DEA to HD, an isotopologue of $H_2$. In this isotopologue, both $H^-$ and $D^-$ show identical forward-backward asymmetry in the angular distribution. We explain these findings using the interference between two quantum paths, with the permanent dipole moment of the asymmetric mass playing no role in the process.

**Introduction:**

Low energy electron attachment to molecules and the evolution of the resultant short-lived molecular negative ion state is a subject of continuing interest both from fundamental physics and a variety of practical applications points of view [1]. The development of momentum imaging techniques for negative ions formed by dissociative electron attachment (DEA) [2] led to several reports on the dynamics of the evolution of these transient states in a wide spectrum of molecules with unprecedented details [1, 3]. Recent advances in theoretical techniques have also helped in understanding the dynamics of simpler molecules [4-6]. The negative ion resonances in molecules, resulting from electron-molecule collisions, act as an effective gateway for transferring kinetic energy into chemical energy in a medium. It has been an important aspect of electron-induced chemistry used for several practical applications [1, 7]. This aspect of electron-induced chemistry through the energy-dependent dynamics and its role in the chemical transformation was found to have an increased potential for chemical control through functional group-dependent site-selective fragmentation of organic molecules by DEA [8]. A recent and rather surprising observation has been the coherent superposition of two resonant states in electron attachment to molecular hydrogen [9], the simplest and most abundant molecule in nature. In fact, it has been shown that such a coherent superposition of two excited states can be accessed in the most general electron scattering channel, namely inelastic scattering channels [10]. DEA forms an important subset of these channels. These findings raise the possibility of a new avenue for chemical control using electron collisions. In the DEA channel, the quantum coherence is observed in the form of an unexpected forward-backward asymmetry in the differential cross-section for the $H^-$ ion generation from the inversion symmetric $H_2$ molecule at 14 eV electron energy [9]. This process also shows an isotope effect where the extent of the forward-backward asymmetry varies in $D_2$ compared to $H_2$ due to longer dissociation time. In that case, how does this quantum coherence play out in the mass-asymmetric isotopologue HD?

The $H^-$ ion yield curve from DEA to $H_2$ shows three peaks at 4eV, 10eV, and 14eV electron energy [11]. The 4eV peak arises due to the dissociation of the lowest attractive $X^2\Sigma_u^+$ anion ground state accessed by the electron capture. This dissociation results in H($^2S$) and $H^-$($^1S$) fragments [12]. The broad peak extending from 6eV to 13eV with a maximum at 10eV has been identified due to the repulsive $B^2\Sigma_g^+$ anion state [13, 14] that dissociates to the lowest limit of H ($^2S$) + $H^-$($^1S$). This peak also shows a contribution from the predissociation of the high-lying bound resonance ($C^2\Sigma_g^+$) [15]. The 14eV peak is found to be due to the coherent excitation of both $^2\Sigma_g^+$ and $^2\Sigma_u^+$ anion states that dissociate to H(n=2) and $H^-$($^1S$) [9]. This coherent superposition leading to the quantum interference of two dissociating paths culminates into inversion symmetry breaking in the form of forward-backward asymmetry in the $H^-$ angular distribution about the incoming electron.

As the DEA process competes with the auto-detachment of the electron from the negative ion resonance formed by the electron attachment, the isotope substitution in the molecule greatly affects the DEA cross-section. This is due to the exponential dependence of the survival probability of the parent anion against auto-detachment on the dissociation time. The three resonances observed in DEA to $H_2$ clearly show this effect in terms of dramatically low cross-sections for heavier isotopologues [11]. However, the most intriguing effect of the isotopic masses on DEA is observed for the 14eV peak in terms of the variation of the forward-backward asymmetry observed [9]. As mentioned earlier, the angular distribution of the $H^-$ ions at the 14eV peak shows a forward-backward asymmetry about the direction of the incoming electron beam. It is an unexpected result from a homonuclear diatomic molecule that possesses an inversion symmetry. The observed asymmetry results from the interference of two dissociating quantum paths arising from the attachment of a single electron and ending in the same dissociation limit ($H(n=2) + H^-(^1S)$). The electron attachment leads to the formation of a coherent superposition of two resonances of opposite parity. The extent of asymmetry is determined by the phase difference between the two paths and the contribution of each of the paths to the dissociation signal. The former depends on the potential energy curve of the involved resonances, and the latter depends on the lifetime of these resonances against the autodetachment. The isotope effect manifests in the form of a change in phase difference due to longer dissociation time for the heavier isotope. Additionally, for the heavier isotope, the amplitude in each path would vary depending on the corresponding widths of the involved resonant states. As for the short-lived resonance, the corresponding path would contribute less to the interference, affecting its contrast, which is the forward-backward asymmetry. This effect is seen in the forward-backward asymmetry in the DEA to $H_2$ and $D_2$ at 14eV electron energy [9].

HD is the hetero-nuclear isotopologue of $H_2$ due to its asymmetric mass. Although the observed forward-backward asymmetry in DEA is due to the interference between the two quantum paths of opposite parity, in principle, no such inversion symmetry is expected to exist in HD as the two atoms have different masses, making the center of mass of the system to differ from the geometric center of the system. Under the Born-Oppenheimer approximation, HD is known to behave like a homonuclear diatomic molecule possessing an inversion symmetry at shorter inter-nuclear separations [16]. However, with the additional terms in Hamiltonian, which couple nuclear motion with electronic motion, the two atoms stand distinguishable due to the different reduced masses of electrons around them. This is particularly observed at higher internuclear separations. Here, HD is expected to behave like a heteronuclear diatom and hence possesses no inversion symmetry [16]. HD is known to possess a permanent dipole moment and, thus, shows a rovibrational spectrum [17], which is used to establish the precise measure of the ratio of the electron and proton mass [18] and limits on quantum forces from dark matter [19]. The difference in the reduced mass of an electron in H and D atoms results in their different excited state spectra, ionization potentials as well as electron affinities. The dissociation limits H(n=2) + D¯ and D(n=2) + H¯ have an energy difference arising from the difference between the energy levels H(n=2) and D(n=2) and the electron affinities of H and D atoms in their ground states. This difference in energy is estimated to be 3 meV [20]. This implies that the two dissociation channels contributing to the 14eV peak in DEA to HD, namely, H¯ and D¯ are distinct.

It is important to understand the effect of this energy difference on the quantum interference observed in the DEA. Xu and Fabrikant found a negligible effect of the small permanent electric dipole moment of HD on the resonance parameters for the ground anion state formed in the electron scattering [21]. The DEA measurements on HD around 10eV electron energy also show that the permanent electric dipole moment of HD does not play any significant role in the electron attachment process [14]. Although the 10eV resonance dissociates to the H($^2S$) + H¯($^1S$) limit, this limit differs in energy for the H¯ and D¯ channels. These dissociation channels show an angular distribution identical to that from the homonuclear isotopologues, highlighting the lack of any significant effect of the permanent dipole moment on the electron attachment [14]. How do this distinguishability in the two channels and the lack of any effect of permanent electric dipole moment on the electron attachment manifest in the quantum interference observed in DEA at 14eV? We have conducted experiments using negative ion momentum imaging to address these questions.

**Experimental set-up:**

The measurements are carried out using a velocity slice imaging (VSI) spectrometer. A magnetically collimated pulsed electron beam (100ns pulse-width) is made to interact at a right angle with an effusive molecular beam produced by a capillary array at room temperature. The anions formed in the interaction region are then extracted into the VSI spectrometer using a pulsed electric field (50V amplitude and 1μs width) with a delay of 100ns after the electron pulse. The details of the spectrometer are reported earlier [22]. The magnetic field used for collimating the electron beam is produced by a pair of Helmholtz coils mounted outside the vacuum chamber. The ions are detected using a two-dimensional position-sensitive detector consisting of a pair of microchannel plates stacked in the chevron geometry, followed by a phosphor screen. The image on the phosphor screen is recorded using a charge-coupled device (CCD) camera. Velocity slice images are taken by pulsing the bias of the detector. The pulsing of the detector corresponds to the arrival of the central slice of the Newton sphere of the relevant ions. In the present experiment, the width of the biasing pulse of the detector is 80ns. The electron energy is calibrated using the H¯ ion-yield curve from $H_2$ as well as using the O¯ ion-yield curve from CO.

**Results and Discussions:**

We first carried out fresh measurements of H⁻ and D⁻ velocity slice images from H$_2$ and D$_2$, respectively, around 14eV. The images obtained were consistent with the earlier reports [23]. Subsequently, we have recorded the H⁻ and D⁻ velocity slice images produced from DEA to HD at various electron energies across the 14eV peak. The images obtained are shown in Figure 1. As can be seen from the figure, both channels show a forward-backward asymmetry where the ion counts are peaking in the backward direction with respect to the electron beam. The two ions show identical momentum values, as expected for the ions generated from HD. However, the D⁻ channel shows relatively poorer momentum resolution. This is because the D⁻ ions with the identical linear momentum value as that of H⁻ would have a smaller spatial spread on the detector due to their heavier mass and, hence, lower speed. The overall angular distribution appears similar to that observed for H⁻ from H$_2$. As expected, the momentum images of both ions show identical sizes. We estimated the forward-backward asymmetry observed in the momentum images using the equation

$$\eta = \frac{I_F - I_B}{I_F + I_B} \tag{1}$$

where $I_F$ and $I_B$ are the intensity of the ion signal in the forward and backward direction of the incoming electron beam. These are determined by integrating the ion counts from the momentum images with positive and negative $p_y$ momentum values, respectively.

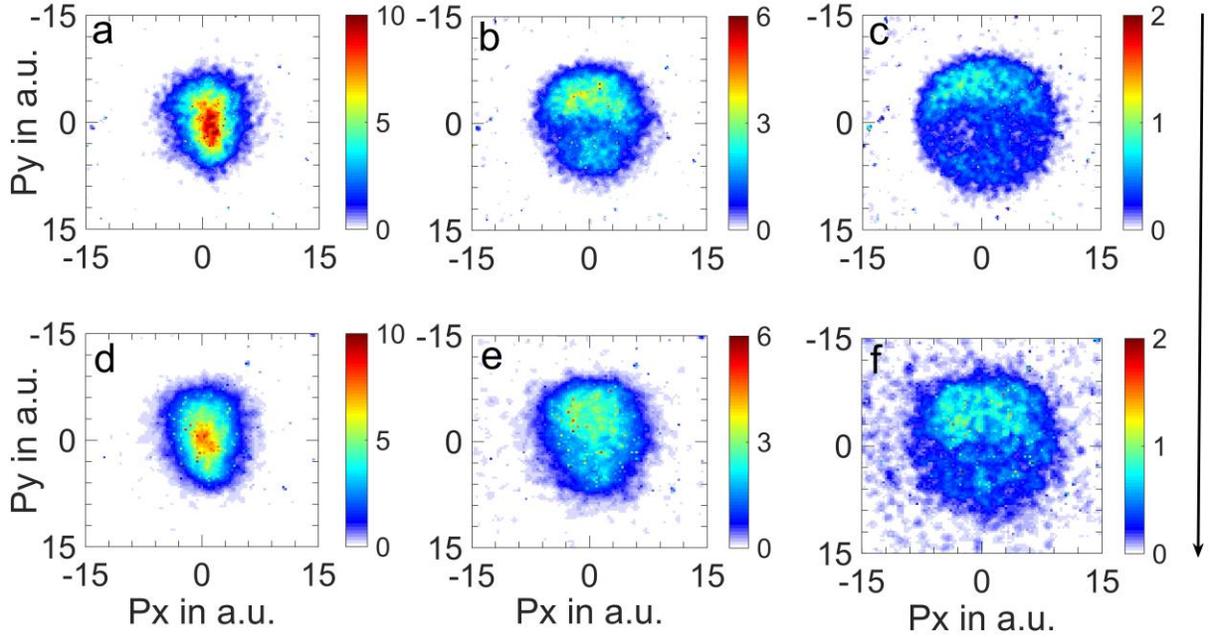

Figure 1: Velocity slice images of H⁻ from HD at (a) 14eV, (b) 14.5eV, and (c) 15eV electron energy and that for D⁻ from HD at (d) 14eV, (e) 14.5eV and (f) 15eV electron energy. The electron beam direction is from top to bottom, as shown in the figure.

The electron energy spread in the experiment was about 0.8 eV FWHM. This shows up as a spread in the ion momentum images. Moreover, the 80ns slice width in time is quite thick for the overall time of flight spread of around 200ns for the H⁻ ions. We estimated the forward-backward asymmetry by looking at the thin ring in the momentum image around its outer edge, as shown in Figure 2(a). The calculated values of the forward-backward asymmetry are given in Table 1 for various electron energies. The use of a thin ring in the momentum image around its outer edge reduces the uncertainty in the obtained results due to electron energy spread. As this image is for the diatomic molecule, the kinetic energy of the fragment has a well-defined relationship with the attaching electron energy, as the system does not have any internal degrees of freedom. This also reduces the effect of uncertainty in the electron energy calibration on the dependence of the asymmetry observed with electron energy. The fresh measurements on $H_2$ and $D_2$ that we carried out, along with the HD measurements, are important to make the appropriate comparison of the forward-backward asymmetry observed in the system. The values obtained for all three systems are given in Table 1.

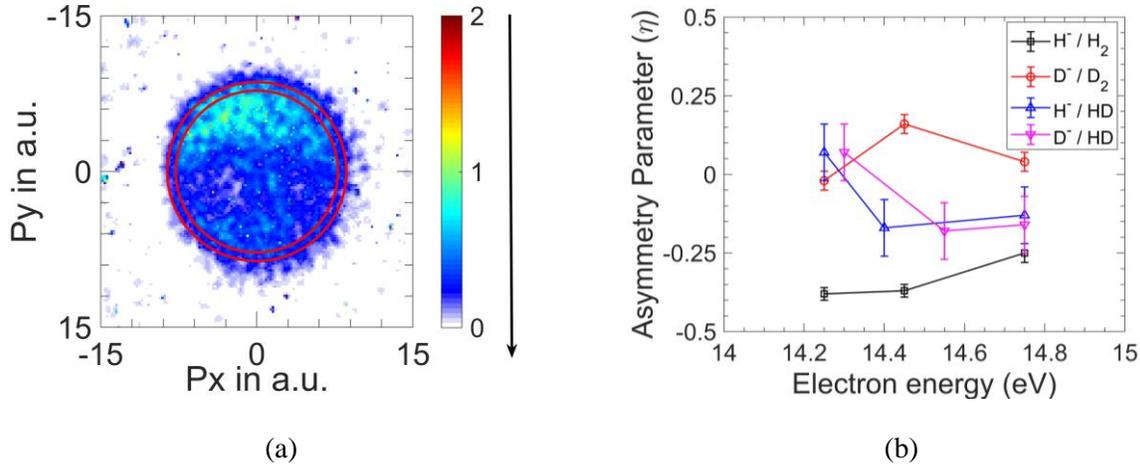

(a) (b)

Figure 2: (a) Velocity slice image of H⁻ from HD at 14.8 eV of electron energy. The annular region shown in the image is used for determining the forward-backward asymmetry to reduce the effect of the finite spread in the electron beam energy. The electron beam direction is from top to bottom as shown (b) Forward-backward asymmetry observed in DEA to $H_2$ (-□-), $D_2$ (-○-), and H⁻ (-Δ-) and D⁻ channels (-∇-) from HD at various electron energies.

As can be seen from Table 1, the forward-backward asymmetry observed in both H⁻ and D⁻ channels from HD is nearly identical within the experimental uncertainty. It is also distinctly different from that observed from $H_2$ and $D_2$, as can be seen in Figure 2(b). The angular distribution of H⁻ and D⁻ formed from *DEA* to $H_2$ and $D_2$ have already been reported for 14 eV resonances by Krishnakumar *et al.* [9]. Our present measurements show that the overall trend of the angular distributions of the anionic fragments is consistent with the previous report [9]. We may add that, as in the case of HD images, we have used a thin ring in the momentum image around its outer edge to reduce the uncertainty due to electron energy spread. This is an improvement we have made in our measurements of the forward-backward asymmetry for $H_2$ and $D_2$ compared to the previous measurements [9].

**Table 1** Forward-backward asymmetry parameter (η) measured for $H^-$ and $D^-$ ions from HD and those obtained for $H^-$ and $D^-$ from $H_2$ and $D_2$, respectively, in the present measurements. The corresponding electron energy obtained from the momentum images is given in brackets.

| Electron energy (eV) | Forward-backward asymmetry (η) | | | |
|---|---|---|---|---|
| | $H^-$ / HD | $D^-$ / HD | $H^-$ / $H_2$ | $D^-$ / $D_2$ |
| 14 eV | 0.07 ± 0.09 (14.25eV) | 0.07 ± 0.09 (14.30eV) | -0.38 ± 0.02 (14.25eV) | -0.02 ± 0.03 (14.25eV) |
| 14.5 eV | -0.17 ± 0.09 (14.40eV) | -0.18 ± 0.09 (14.55eV) | -0.37 ± 0.02 (14.45eV) | 0.16 ± 0.03 (14.45eV) |
| 15 eV | -0.13 ± 0.09 (14.75eV) | -0.16 ± 0.09 (14.75eV) | -0.25 ± 0.03 (14.75eV) | 0.04 ± 0.03 (14.75eV) |

To understand the forward-backward asymmetry observed in the homonuclear diatom $H_2$ and $D_2$, the coherent superposition of two resonances of opposite parities was considered [9]. Based on the overall angular distribution observed for DEA, it was estimated that the contributing resonances would be of the terms $^2\Sigma_g^+$ and $^2\Sigma_u^+$ where they overlap in the Frank-Condon region of the neutral ground state $^1\Sigma_g^+$. The lifetime against auto-detachment for the $^2\Sigma_g^+$ state involved has been reported to be 9 fs [24], and the lifetime for the $^2\Sigma_u^+$ state was taken as a parameter to estimate the forward-backward asymmetry observable from $H_2$ and $D_2$. Interestingly, the momentum images observed for both $H^-$ and $D^-$ channels from HD are closely similar and resemble those from $H_2$. Due to the asymmetric mass, HD is expected to show no inversion symmetry. This is particularly valid for the larger internuclear separation [16]. As a result, the contributing resonances in the case of either channel, namely $H^-$ and $D^-$, are $^2\Sigma^+$. This can be ascertained using the angular distribution observed in either channel. The corresponding angular distributions, obtained for both channels at 14.5 eV, are shown in Figure 3. The ground state of the neutral molecule is $^1\Sigma^+$. We have fitted the angular distribution using the equation

$$I(k,\theta) = A_s^2 + A_p^2 \cos^2\theta + 2A_s A_p \cos\theta \cos\varphi \qquad (2)$$

where $A_s$ and $A_p$ correspond to the amplitude of the s-wave and p-wave capture by the neutral molecule, $\theta$ is the angle of ejection of anion fragment about the incoming electron beam, and $\varphi$ is the relative phase between the two partial waves captured. This fit indicates that the NIR state leading to either of the anion channels is of $^2\Sigma^+$ symmetry.

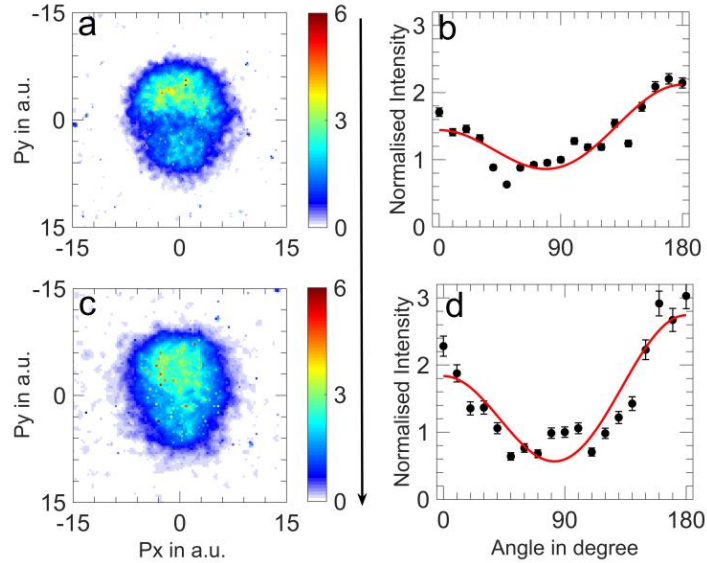

Figure 3: Momentum images and corresponding angular distributions of H⁻ and D⁻ ions from HD at 14.5 eV electron energy. (a) Momentum image of H⁻; (b) H⁻ angular distribution; (c) Momentum image of D⁻; (d) D⁻ angular distribution. The arrow indicates the direction of the electron beam.

However, as discussed earlier, the two dissociation limits are energetically not the same and cannot result from the same negative ion state. Hence, the two ion signals cannot arise from the same NIR state. As the two states of opposite parity contributing to the coherent superposition on electron attachment to the homonuclear counterparts overlap in the Frank-Condon region, any term in the Hamiltonian that lifts the inversion symmetry would mix these two states. The resulting states would independently contribute to the two channels, namely H⁻ + D (n=2) and D⁻ + H (n=2). In that case, the two resulting linear combinations would be

$$\psi_1 = {}^2\Sigma_g^+ + {}^2\Sigma_u^+ \tag{3a}$$

$$\psi_2 = {}^2\Sigma_g^+ - {}^2\Sigma_u^+ \tag{3b}$$

The corresponding angular distribution expected from these states would be given by

$$I_1(k,\theta) = A_{s_1}^2 + A_{p_1}^2 \cos^2\theta + 2A_{s_1}A_{p_1}\cos\theta\,\cos\varphi_1 \tag{4a}$$

$$I_2(k,\theta) = A_{s_2}^2 + A_{p_2}^2 \cos^2\theta - 2A_{s_2}A_{p_2}\cos\theta \cos\varphi_2 \qquad (4b)$$

where the terms with the *s* and *p* subscripts arising from the *s*-wave and *p*-wave capture, and $\varphi_1$ and $\varphi_2$ are the relative phases between the two partial waves for each of the resonances. If the two resulting states are energetically close to each other, the phases $\varphi_1$ and $\varphi_2$ will have almost the same values. In such a scenario, the expected angular distributions would peak in opposing directions for the two channels. In other words, if the ion intensity for H⁻ peaks in the forward direction, then that for D⁻ will peak in the backward direction or vice versa, which is not observed in the present measurements.

We note that the absolute cross-sections of the two channels of HD for 14 eV resonance are more or less identical, which is clear from the momentum images. To understand this result, we have estimated the dissociation time and the survival probability for the two resonances. For this purpose, we have picked up two potential energy curves in this energy range from the literature [25] that run close to each other and lead to the dissociation limit with one of the constituent atoms in the excited state (n=2). As the dissociation limits for both the channels differ by 3meV, we assume that the upper curve dissociates to the D (n=2) + H⁻ channel and the lower state dissociates to H (n=2) + D⁻ channel. The absolute cross-section of the *DEA* process is given by

$$\sigma_{DEA} = \sigma_C \times p(E) \qquad (5)$$

where $\sigma_c$ is the electron capture cross-section and *p(E)* is the survival probability against auto-detachment. The survival probability is given by

$$p(E) = \exp\left(-\frac{t_d}{\tau}\right) \qquad (6)$$

where $t_d$ is the dissociation time, and $\tau$ is the average lifetime of the resonance against auto-detachment. One can estimate the survival probability for the two channels if one knows the auto-detachment lifetime of the resonances. Using the selected potential energy curves, we estimate the dissociation time for each of them for HD to be 14.43 fs and 14.06 fs for 15 eV incident electron energy.

In order to have the same absolute cross-section for both the channels, either the capture cross-section and survival probability for each resonance are identical, or the survival probabilities are inversely proportional to the electron capture cross-sections. Here, we assume that the two resonances have identical capture cross-sections. For identical values of the survival probabilities, the two resonant states must have very similar lifetimes, as their dissociation times are pretty close. This is possible because the two resonances are mixtures of two corresponding resonances observed in the homonuclear isotopologue with opposite parity. As these resonant states are not explicitly calculated in the past, we cannot estimate their lifetimes.

From the 10 eV measurements for HD [14], it is observed that the permanent dipole moment arising from the asymmetric mass of HD does not play any significant role in the electron capture process. This observation implies that HD behaves like a homonuclear diatom for the electron capture process, and similar to $H_2$ and $D_2$, the 14 eV resonance corresponds to the linear superposition of the two resonances of opposite parity. However, our earlier analysis also rules out the two-state description of two participating resonances at larger inter-nuclear separations, leading to two distinct dissociation limits.

De Lange *et al.* [16] have developed a procedure to construct long-range potential near the *n=2* dissociation limit, which results in the breakdown of *g-u* symmetry. In $H_2$ and $D_2$, due to the identical atoms, the *n=2* dissociation limit has a two-fold degeneracy. But in the case of HD, this degeneracy is lifted. Therefore, for HD, there are two groups of limits for *n=2* dissociation, H(*2l*) + D(*1s*) and H(*1s*) + D(*2l*), and each limit has a two-fold near degeneracy for *l=0* or *l=1*. They have pointed out that the potential energy curves for isotopologues of $H_2$ coincide for small internuclear separations. However, for larger internuclear distances, the potential energy curves for HD deviate from the adiabatic curve and converge to one of the dissociation limits without crossing each other. Implementing this description implies that the electron capture process would be identical for $H_2$, $D_2$, and HD in the Frank-Condon region.

At the dissociation limit of the 14 eV resonance, the fragmented negative ion is in the ground state, whereas the neutral atom is in the first excited 2s or 2p state. If the excited atom is in the 2*s* state, the possible molecular states would be $^2\Sigma_g^+$ and $^2\Sigma_u^+$, and if the exited atom is in the 2*p* state, then the possible molecular states would be $^2\Sigma_g^+$, $^2\Sigma_u^+$, $^2\Pi_g$ and $^2\Pi_u$. For homonuclear diatomic molecular anions $H_2^-$ and $D_2^-$, all the states mentioned above converge, as the 2s and 2p are the near degenerate states. As explained before, the forward-backward asymmetry is due to the interference of two dissociating quantum paths. These paths involve resonances of $^2\Sigma_g^+$ and $^2\Sigma_u^+$ symmetry coherently accessed by the transfer of the *s*-wave and *p*-wave, respectively, in the electron attachment process.

For heteronuclear counterpart, $HD^-$, *n=2* dissociation limit has two distinct channels H(*2l*) + $D^-$ ($^1S$) and D(*2l*) + $H^-$ ($^1S$). The excitation is localized to either of the distinct H or D atoms in each channel. These two channels are separated by 3 meV of energy, about two orders of magnitude higher than the fine structure splits [20]. The states for the heteronuclear molecule would be a linear combination of the homonuclear molecular states. The linear combination of the four states would give four states of $HD^-$ and these mutually orthogonal combinations would be as follows:

$$1\,^2\Sigma_g^+ + 1\,^2\Sigma_u^+ + 2\,^2\Sigma_g^+ + 2\,^2\Sigma_u^+ \tag{7a}$$

$$1\,^2\Sigma_g^+ + 1\,^2\Sigma_u^+ - 2\,^2\Sigma_g^+ - 2\,^2\Sigma_u^+ \tag{7b}$$

$$1\,^2\Sigma_g^+ - 1\,^2\Sigma_u^+ + 2\,^2\Sigma_g^+ - 2\,^2\Sigma_u^+ \tag{7c}$$

$$1\,^2\Sigma_g^+ - 1\,^2\Sigma_u^+ - 2\,^2\Sigma_g^+ + 2\,^2\Sigma_u^+ \tag{7d}$$

The prefix 1 refers to those $\Sigma$ states involved in the 14 eV *DEA* channel contributing to the forward-backward asymmetry in H$_2$ and D$_2$, and the prefix 2 refers to those $\Sigma$ states which are not responsible for the 14 eV *DEA* peak. Among these four combinations, two would converge to the H(*2l*) + D$^-$($^1S$) limit, and the remaining two would converge to the D(*2l*) + H$^-$($^1S$) limit. Here, based on the experimental observation of the identical angular distributions for the two channels, we predict that the combinations converge in pairs either like (7a) & (7c) and (7b) & (7d) or (7a) & (7d) and (7b) & (7c). A detailed theoretical calculation is required to identify these anionic molecular states. This picture explains the observed forward-backward asymmetry for all possible isotopologues.

Charron *et al.* [26] theoretically demonstrated the coherent control of isotope separation in the photodissociation of HD$^+$ molecules. Applying a coherent addition of the fundamental radiation with its second harmonic can lead to asymmetries in fragment angular distribution. Their simulation shows that the H$^+$ and D$^+$ ions are ejected preferentially in the forward and backward directions, respectively, about the resultant asymmetric light field formed by the zero phase difference between the two colours. On the contrary, B. Sheehy *et al.* [27] experimentally observed that in the two-colour photodissociation of HD$^+$, both the fragment ions (H$^+$ and D$^+$) show identical asymmetry. This asymmetry varies with the relative phase between the two light fields but remains similar for both channels. Charron *et al.* clarified this difference in the photodissociation pattern as follows [26]. There are two distinct regions of the enhanced probability of photodissociation: low-frequency and high-frequency regimes of light. In the low-frequency regime, the permanent dipole moment of HD plays its role in controlling the dissociation pattern in which the H$^+$ and D$^+$ ions are preferentially ejected in the opposite direction. On the contrary, in the high-frequency regime, the effect of permanent dipole moment is significantly reduced, due to which the dynamics of HD$^+$ are very similar to H$_2^+$, and no appreciable difference between the H$^+$ and D$^+$ partial probabilities can be induced in this regime as observed experimentally by B. Sheehy *et al.* Therefore, it is concluded that the preferential ejection of H$^+$ and D$^+$ ions in opposite directions in the dissociation of HD$^+$ is due to its permanent dipole moment. The results obtained for the *DEA* to HD that results from the coherent superposition of the two resonances also show identical behaviour. This also shows that the permanent dipole moment of the HD due to asymmetric mass does not play any role in the quantum interference, which is in accordance with the photodissociation case.

## Conclusion:

To conclude, we have shown that DEA to HD shows near identical forward-backward asymmetry in both H⁻ and D⁻ channels. The two channels have equal *DEA* cross-sections. We have explained the observed identical asymmetry in both channels using the coherent superposition of two anion states of opposite parities, as presented earlier for $H_2$ and $D_2$. However, the distinct dissociation limits for the two channels are described by the linear combinations of four anion states, two of which participate in the DEA process at 14 eV. As the two participating states have the same signs in the dissociation limit description, the forward-backward asymmetry observed in the two channels is identical. The observation also supports the earlier findings that the permanent electric dipole of the mass asymmetric HD does not play any significant role in the electron capture process and, hence, in the symmetry breaking. These results also indicate a dire need for reliable calculations using modern theoretical tools for determining the anion resonances in these simplest diatomic molecules in terms of their potential energy curves and their lifetimes.

## Acknowledgement:

E.K. acknowledges the Raja Ramanna Fellowship from the Department of Atomic Energy, India. S.S., A.K. and V.S.P. acknowledge the financial support from the Department of Atomic Energy, India under Project Identification No. RTI4002.